\documentclass[prc,reprint,superscriptaddress,amsmath,amssymb,floatfix,showpacs,notitlepage,nofootinbib,preprintnumbers]{revtex4-1}

\usepackage{dcolumn}
\usepackage{graphicx}
\usepackage{hyperref} 
\usepackage[usenames]{xcolor}

\newcommand{\p}[1]{\protect{(#1)}}
\newcommand{\pfrac}[2]{\left(\frac{#1}{#2}\right)}
\newcommand{\rmi}{\mathrm{i}}
\newcommand{\rmd}{\mathrm{d}}
\newcommand{\avg}[1]{\left\langle#1\right\rangle}

\newcommand{\caT}{\mathcal{T}}

\newcommand{\bfK}{\mathbf{K}}
\newcommand{\bfR}{\mathbf{R}}
\newcommand{\bfk}{\mathbf{k}}
\newcommand{\bfq}{\mathbf{q}}

\newcommand{\bfu}{\mathbf{u}}
\newcommand{\bfx}{\mathbf{x}}
\newcommand{\sfD}{\mathsf{D}}
\newcommand{\hbx}{\boldsymbol{\hat{\mathbf{x}}}}
\newcommand{\hby}{\boldsymbol{\hat{\mathbf{y}}}}
\newcommand{\hbz}{\boldsymbol{\hat{\mathbf{z}}}}
\newcommand{\bhe}{\boldsymbol{\hat{\mathbf{e}}}}
\newcommand{\hbk}{\boldsymbol{\hat{\mathbf{k}}}}
\newcommand{\hbq}{\protect{\boldsymbol{\hat{\mathbf{q}}}}}
\newcommand{\pF}{p_\mathrm{F}}
\newcommand{\eF}{\varepsilon_\mathrm{F}}
\newcommand{\kD}{k_\mathrm{D}}
\newcommand{\kTF}{k_\mathrm{TF}}
\newcommand{\rten}{\rho_{10}}
\newcommand{\melt}{\mathrm{m}}
\newcommand{\wpl}{\Omega_{\rm P}}

\newcommand{\gcc}{\text{g cm}^{-3}}

\newcommand{\K}{\text{K}}
\newcommand{\fermi}{\text{fm}}

\newcommand{\OPA}{\protect{\text{OPA}}}
\newcommand{\CMC}{\protect{\text{CMC}}}
\newcommand{\QMC}{\protect{\text{QMC}}}

\begin{document} 

\preprint{INT-PUB-15-020}
\title{Quantum Monte Carlo calculations of the thermal conductivity of neutron star crusts}

\author{Sajad Abbar}
\affiliation{Department of Physics and Astronomy, University of
  New Mexico, Albuquerque, NM 87109}
\author{Joseph Carlson}
\affiliation{Theoretical Division, Los Alamos National Laboratory,
  Los Alamos, NM 87545} 
\author{Huaiyu Duan
  }
\affiliation{Department of Physics and Astronomy, University of
  New Mexico, Albuquerque, NM 87109}
\author{Sanjay Reddy}
\affiliation{Institute for Nuclear Theory, University of
  Washington, Seattle, WA 98195-1550}

\begin{abstract} 
We use the quantum Monte Carlo (QMC) techniques to calculate the 
static structure function $S(q)$ of a one-component ion lattice and
use it to calculate the  thermal conductivity 
$\kappa$ of high-density solid matter expected in the neutron star
crust. By making detailed comparisons with the results for the thermal conductivity 
obtained using standard techniques based on the one-phonon approximation (OPA) valid at low temperature, and the multi-phonon 
harmonic approximation expected to be valid over a wide range of temperatures, 
we asses the temperature  regime where $S(q)$ from QMC can be used directly 
to calculate  $\kappa$. We also compare the QMC results to those obtained using classical Monte Carlo to quantitatively 
asses the magnitude of the quantum corrections.  We find that quantum effects became relevant 
for the calculation of $\kappa$ at temperature $T \lesssim  0.3 ~\wpl$, 
where $\wpl$ is the ion plasma frequency. At $T \simeq  0.1 ~\wpl$
the quantum effects suppress $\kappa$ by about $30\%$.  The comparison with the results of the OPA indicates that dynamical 
information beyond the  static structure is needed when $T \lesssim  0.1~ \wpl$. These quantitative comparisons 
help to establish QMC as a viable technique to calculate $\kappa$ at moderate temperatures in the range 
$T=0.1-1~\wpl$ of relevance  to the study of accreting neutron stars. This finding is especially important because 
QMC is the only viable technique so far for calculating $\kappa$ in multi-component systems at low-temperatures. 
\end{abstract} 


\maketitle 



\section{Introduction}
         \label{sec:Intro}
Observations of transient phenomena in accreting neutrons stars and
magnetars \cite{Eichler:1989,Rutledge:2002} have motivated recent
attempts to model the thermal evolution of the outermost regions of
the neutron star called the crust
\cite{Shternin:2007,BrownCumming:2009,PageReddy:2012,PageReddy:2013}. These
studies have shown that the thermal conductivity of the crust plays a
very important role in shaping temporal structure of x-ray emission
from these systems. In this study we revisit the calculation of  the
thermal conductivity of the outer crust of the neutron star, where the
typical densities are in the range of $ 10^{8}-10^{11}\,\gcc$ and
temperatures are expected to be in the range of $10^7-10^9$ K. Under
these conditions nuclei are pressure ionized and form a solid phase at
low temperature. In contrast, electrons are relativistic, weakly
coupled and very degenerate. For these reasons electrons dominate
the thermal conductivity of the neutron star crust.  

Electron conduction in the neutron star crust is limited by
electron-ion scattering. Ions in the crust have large atomic number
($25\lesssim Z \lesssim40$),
and ionic structure and dynamics are strongly correlated by
Coulomb interactions. Consequently, the amplitude for electron
scattering off different ions interfere. Accounting for such
interference under arbitrary ambient conditions in a multi-component
plasma (MCP) is a challenging many-body problem. At low temperature and for
the case when the ground state is a simple solid containing only one
ion species, pioneering works by Flowers and Itoh
\cite{FlowersItoh:1976} and by Yakovlev and Urpin
\cite{YakovlevUrpin:1980} have shown that the one-phonon 
approximation (OPA) adequately describes the collective response of 
the ions which is needed to calculate the electron scattering rate.  More recent
studies have
greatly improved the description of electron scattering in a one
component plasma (OCP) by including the effects of the multi-phonon
processes with the harmonic  
approximation \cite{BaikoEtal:1998,Potekhin:1999}. It is now possible to 
calculate the electron thermal conductivity of an OCP over the full
range of temperatures of interest to neutron star astrophysics.  For
MCP, classical molecular dynamics has proven
to be useful and is expected to provide a quantitive description of
the  thermal conductivity in the high temperature limit when $T \gtrsim
\wpl$ \cite{HorowitzCaballeroBerry:2009}. However, neither of these
methods are suitable to describe electron scattering in
multi-component systems at low temperature. This observation is the
primary motivation for our study here  
where we establish that Quantum Monte Carlo (QMC) methods, which are
simply generalizable to MCP,  
can be used to calculate the thermal conductivity at the relatively
low temperatures of interest in astrophysics.   

Quantum Monte Carlo (QMC) methods have been successful in determining
the ground state properties of diverse strongly correlated
many-particle systems and can be adapted to study MCPs at low
temperature when quantum effects become relevant. In this study, which
is a first step towards developing a QMC approach to MCPs, we use the
QMC method to study the ion-ion correlations in an OCP at low
temperature and calculate the thermal conductivity.  We also use the
dynamic matrix to calculate the detailed phonon spectrum at low
temperature which is in turn used to
calculate $\kappa$. These calculations allow us to make
quantitative comparisons between the results obtained with the Monte
Carlo techniques and those obtained with various approximations
used in the literature. These comparisons, allow us to reassess 
various theoretical approaches in some detail and 
provide useful new insights about the role of quantum effects, the low-energy 
strength of the ion response function, Bragg scattering  and
multi-phonon effects.

The rest of the paper is organized as
follows. In 
Section~\ref{sec:kappa} we review the basic properties of an OCP and
define the relationship between the electron conductivity and the ion
dynamical structure function $S(\omega,\bfq)$. In
Section~\ref{sec:phonon} we discuss 
the phonon spectrum and use it to calculate $S(\omega,\bfq)$ in
OPA. In Section~\ref{sec:mc} we describe and use 
both classical and quantum Monte Carlo methods to calculate the static
structure function $S(q)$ and compare these results to those obtained
in OPA. In
Section~\ref{sec:results} we 
calculate $\kappa$ using different approaches and approximations, and
we identify the
temperature regimes where these approximations are valid.  We also
outline the strategies to extend QMC calculations to MCPs in future work.
Throughout this paper we adopt the natural physical units with
$\hbar=c=k_\text{B}=1$.

\section{Thermal conductivity of the outer crust}
\label{sec:kappa}

In the simplest scenario the outer crust at given density is composed of cold
catalyzed matter of a single ion species. The mass number $A$ and
charge $Z$ of the ions are density dependent and are determined by
minimizing the total energy of the system.  The ground state of such matter is a
strongly correlated OCP with bare nuclei immersed in a degenerate and
weakly coupled electron gas. The characteristic energy of the electron is
set by its Fermi momentum  
\begin{align}
  \pF &= (3\pi^2 n_e)^{1/3}
  \nonumber\\
  &\approx (25\,\fermi)^{-1}\, \pfrac{Z}{30}^{1/3}
  \pfrac{A}{80}^{-1/3}\rten^{1/3}, 
\end{align}
where $n_e$ is the number density of electrons,
and $\rten$ is the mass density in units of  $10^{10}\,\gcc$. 
For the densities of interest to our study, which
are typically in the range of $10^8-10^{11}~\gcc$, $\pF$ is much larger
than electron mass $m_e$, and it is a good
approximation to treat electrons as ultra-relativistic. In
contrast, ions are heavy and correlated. One characteristic
energy is set by the ion plasma frequency  
\begin{align}
  \wpl &= \left(\frac{4\pi Z^2 e^2 n_I}{M}\right)^{1/2}
  \nonumber\\
  &\approx (2.9\times10^8\,\K) \pfrac{Z}{30} \pfrac{A}{80}^{-1} \rten^{1/2},
\end{align}
where $n_I=n_e/Z$ is the ion density, $M\approx A m_p$ is the mass of
the ion with $m_p$ being the proton mass, and $e^2\approx1/137$ is the
fine structure constant in natural units. 

The typical Coulomb energy is of order $Z^2 e^2/a$,
where 
\begin{align}
 a=
\left(\frac{4\pi n_I}{3 }\right)^{-1/3} 
\approx (147\,\fermi)\, \pfrac{A}{80}^{1/3} \rten^{-1/3}
\label{eq:a} 
\end{align}
is the inter-ion distance. The temperature $T$ provides a measure of
the ``extractable'' kinetic energy or the thermal energy of the ions.
The ratio between the Coulomb energy and the thermal energy of the ions 
is a measure of the importance of interactions in the plasma and
defines the dimensionless Coulomb parameter
\begin{align} 
\Gamma = \frac{Z^2 e^2}{a T}.
\end{align}
In a weakly coupled plasma $ \Gamma \ll 1$, and Coulomb interaction
can be studied within perturbation theory. Numerical simulations of
the OCP have shown that ions crystallize into a solid state when
$\Gamma > \Gamma_\melt \approx 175 $
\cite{SlatteryDoolenDeWitt:1980,SlatteryDoolenDeWitt:1982,JonesCeperley:1996}.
The melting temperature of the solid  
\begin{align}
T_\melt =\frac{Z^2 e^2}{a \Gamma_\melt}
\approx 2.0\, \wpl \pfrac{Z}{30} \pfrac{A}{80}^{2/3} \rten^{-1/6}
\end{align}
can be correspondingly defined. We note that electron screening
modifies the Coulomb potential generated by the ion  at large
distances. The modified potential is
\begin{align}
V(r) = \frac{Z e }{r}\, e^{- r\,\kTF},
\end{align}
where 
\begin{align}
\kTF &= \left(4\pi e^2 \frac{\partial n_e}{\partial
  \mu_e}\right)^{1/2}
\nonumber\\
&\xrightarrow{\pF \gg m_e}  \sqrt{\frac{4e^2}{\pi}}~\pF
\approx( 1.7a)^{-1} \pfrac{Z}{30}^{1/3}
\label{eq:kTF}
\end{align}
is the Thomas-Fermi (screening) momentum.
Because $\kTF a < 1 $ for the densities of interest, screening will
not greatly alter the nearest neighbor ion-ion interaction, and the
Coulomb parameter of the OCP without screening continues to provide a
reasonable measure of the strength of interactions and the melting
temperature.  In this paper we have chosen to present results at
fiducial densities $10^{10}$ and $10^{11}\,\gcc$, labelled as LD and HD,
respectively. The chemical compositions in these two cases are chosen
according to \citet{PageReddy:2012} for catalyzed matter in the outer
crust. We list  
the physical conditions of the two cases in Table~\ref{tab:cases}. 

\begin{table*}[bht]
\caption{Key parameters for cold catalyzed matter in neutron star
  crust at two fiducial densities. The compositions are chosen
  according to \citet{PageReddy:2012}. } 
\centering
\begin{ruledtabular}
\begin{tabular}{c c c c d  d d d d d} 
  Name & Lattice  & $\rho\,(\gcc)$ & Ion &
  \multicolumn{1}{c}{$a\,(\fermi)$} &
  \multicolumn{1}{c}{$\wpl\,(10^9\,\K)$} &
  \multicolumn{1}{c}{$T_\melt\,(10^9\,\K)$} &
  \multicolumn{1}{c}{$\pF^{-1}\,(\fermi)$} &
  \multicolumn{1}{c}{$\kTF^{-1}\,(\fermi)$} &
  \multicolumn{1}{c}{$\kD^{-1}\,(\fermi)$} \\
  \hline 
  LD & BCC & $10^{10}$ & $^{84}_{34}$Se & 149 & 0.32 &  0.75 & 24 &
  249 & 62\\
  HD & BCC &  $10^{11}$ & $^{80}_{28}$Ni & 68 & 0.87 &  1.11 & 12 &
  121 & 28\\
\end{tabular}
\end{ruledtabular}
\label{tab:cases}
\end{table*}

The dynamic structure function in the crystalline state is given by  
\begin{align}
S(\omega,\bfq) &= \frac{1}{N}\sum_{i,j=1}^N
e^{-\rmi\bfq\cdot(\bfR_i-\bfR_j)}
\nonumber\\
&\quad\times\int\frac{\rmd t}{2\pi} e^{\rmi\omega t}
~\langle e^{\rmi\bfq\cdot\bfu_j(0)}\,
e^{-\rmi\bfq\cdot\bfu_i(t)}\rangle_T,
\end{align} 
where $N$ is the total number of ions in the crystal, 
$\bfR_i$ is the equilibrium position of the ion on the $i$'th site,
$\bfu_i(t)$ is the displacement of this ion at time $t$, and
$\langle \cdots \rangle_T$ denotes the thermal average.

The electron thermal conductivity $\kappa$ is mainly limited by
electron-ion scattering and can be written as
\begin{align}
\kappa = \frac{\pi^2 T n_e }{3 \eF \nu_\kappa},
\end{align}
where $\eF\approx\pF$ is the electron Fermi energy, and effective
collision rate is \cite{FlowersItoh:1976}
\begin{align}
\nu_\kappa &= \frac{2}{3}\frac{\eF}{Z (2\pi)^3}
\int_0^{2\pF}\rmd q\, q^3 |v(q)|^2 
S'_\kappa(q)\,.
\label{eq:nu_kappa}
\end{align}
In the above expression 
\begin{align}
|v(q)|^2 = e^2 |V(q)|^2 \left(1-\frac{q^2}{4\pF^2}\right)
\end{align}
is the
square of the scattering matrix element for electron-ion interaction
with momentum exchange $\bfq$,   
and the screened Coulomb potential generated by the ion in momentum space is 
\begin{align} 
V(q) = \frac{4\pi Z e}{\epsilon(q)~q^2}\,, 
\end{align} 
where
\begin{align}
  \epsilon(q)=1+\frac{\kTF^2}{q^2}
\end{align}
is the static dielectric function in
the Thomas-Fermi approximation. The effects due to ion-ion
correlations on electron scattering are included in
Eq.~\eqref{eq:nu_kappa} through  
\begin{align}
S'_\kappa(q)=\int_{-\infty}^\infty\rmd\omega\,  \avg{S'(\omega,\bfq)}_\hbq
w_\kappa(\omega/T, q), 
\label{eq:sprime}
\end{align}
where $S'(\omega,\bfq)$ is the inelastic part of the dynamic structure
function, $\avg{\cdots}_\hbq$ is the average over the direction of unit vector 
$\hbq=\bfq/q$, and
\begin{align}
w_\kappa(z=\omega/T, q)  = \frac{z}{e^{z}-1}
\left[1+\frac{z^2}{\pi^2}
\left(\frac{3\pF^2}{q^2}-\frac{1}{2}\right)\right].
\label{eq:w}
\end{align}
We note that the elastic Bragg scattering 
does not contribute to the conductivity because it has been accounted for
in the ground state which leads to the electronic band structure.

Because the response at high frequency $|\omega|\gg \wpl$ cannot involve 
collective motion of the ions, we expect that most of the strength of the 
dynamic response will reside at energies that are comparable to $\wpl$.   
Therefore, when $T \gtrsim \wpl $ it is a good approximation to 
retain only the leading terms of $w_\kappa(z,q)$ in $z=\omega/T$ in
the integrand in Eq.~\eqref{eq:sprime}, and the static approximation
$S'_\kappa(q)\approx S'(q) $ is valid, where
\begin{align}
  S'(q) = \int_{-\infty}^\infty \rmd\omega\, 
  \avg{S'(\omega, \bfq)}_\hbq,
  \label{eq:sofq}
\end{align}
is the inelastic part of the static structure function.
At very low temperature $T\ll\wpl$, however, the exponential factor
$1/(e^{-\omega/T}-1)$ in $w_\kappa(z,q)$ dominates. In this limit the static
approximation breaks down, 
and $S'_\kappa(q)\ll S'(q) $.
Between these two temperature limits two 
competing factors in $w_\kappa(z,q)$ dominate in different ranges of $q$.
For large-angle scattering (with large $q$ values) the
exponential factor still dominates, and $S'_\kappa(q) < S'(q)$.  But for
small-angle scattering (with small $q$ values) the factor
$\pF^2/q^2$ can dominate, and $S'_\kappa(q) > S'(q)$.
We will determine these temperature limits for OCP using OPA in the
next section. 

\section{Phonon spectrum and the dynamical response}
\label{sec:phonon}

\begin{figure*}[htb] 
\centering
$\begin{array}{@{}cc@{}}
\includegraphics*[scale=0.55]{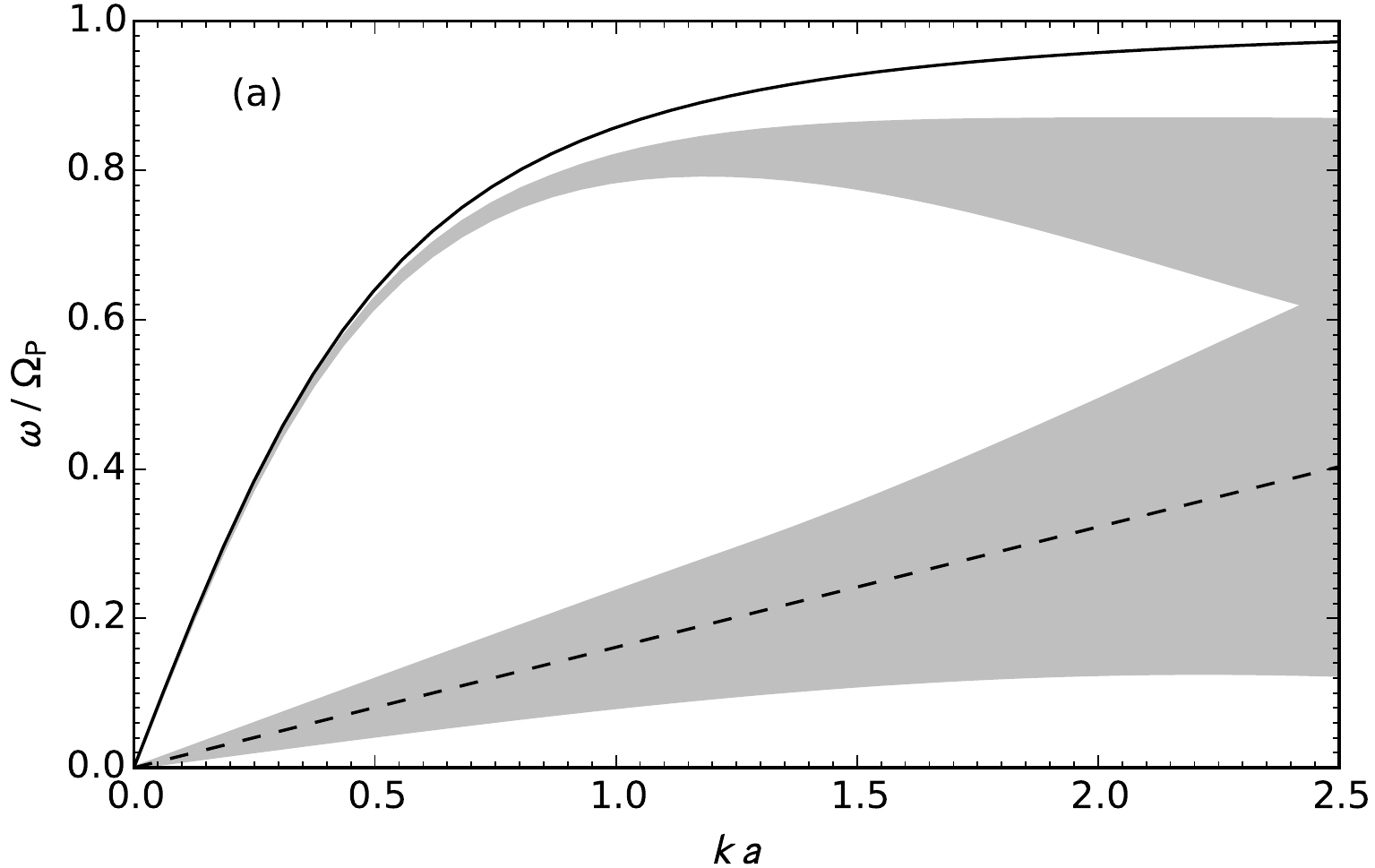}&
\includegraphics*[scale=0.55]{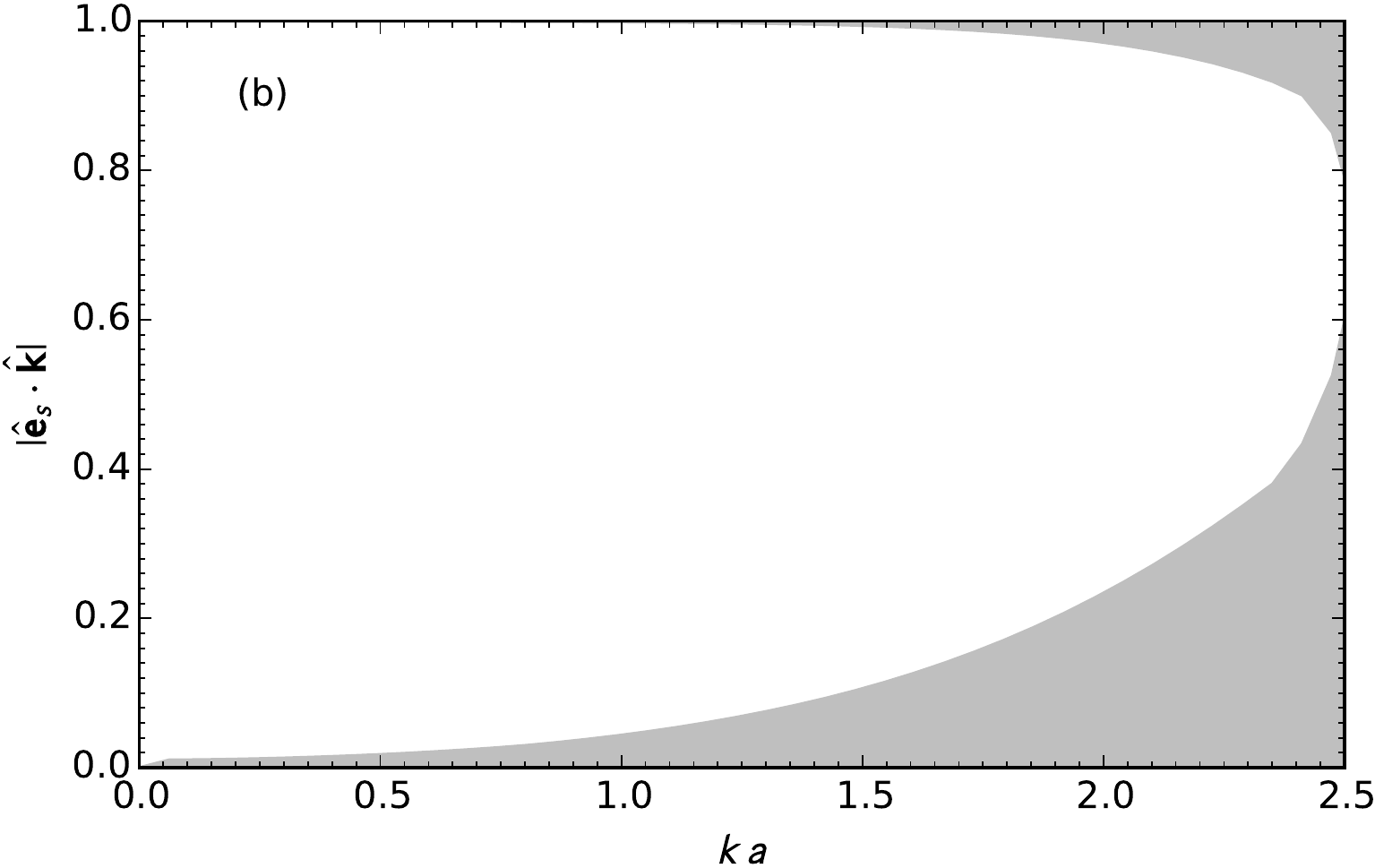}
\end{array}$
\caption{The dispersion relations of the photon modes of the Coulomb
  lattice. Left panel: The phonon frequencies $\omega$ as
  functions of wave number $k$. The dashed and solid curves are for the
  transverse and longitudinal modes described by Eqs.~\eqref{eq:ct} and
  \eqref{eq:omega-l}, respectively. The shaded region gives the
  range of the eigenvalues of the dynamic matrix in Eq.~\eqref{eq:D}.
  Right panel: The shaded region gives the range of
  $|\bhe_s\cdot\hbk|$ for the eigenvectors $\bhe_s(\bfk)$ of
  the dynamic matrix. The results in both panels
  are computed for the LD matter.}
\label{fig:dispersion}
\end{figure*}

At low temperature the characteristic distance scale for ion motion in
lattice is
\begin{align}
\lambda_I &= \left(\frac{1}{2 M \wpl}\right)^{1/2}
\approx (3.2\,\fermi)\, 
\pfrac{Z}{30}^{-1/2}\rten^{-1/4},
\end{align}
which is much shorter than the inter-ion distance $a$.  Under these
conditions the restoring force on the ion is quadratic in the
displacement $\bfu_i(t)$, and the detailed phonon spectrum can be
calculated by using the dynamic matrix \cite{Carr:1961,Ashcroft}
\begin{align}
\sfD(\bfk) 
&=2\sum_{i=1}^N\sin^2\left(\frac{\bfk\cdot\bfR_i}{2}\right)
\left[\frac{\partial^2 V(\bfx)}{\partial\bfx
\partial\bfx^T}\right]_{\bfx=\bfR_i}.
\label{eq:D}
\end{align}
The phonon frequencies
$\omega_s(\bfk)$ ($s=1,2,3$) are obtained by solving the eigenvalue equation
$M\omega_s^2(\bfk)-\sfD(\bfk)=0$, and the corresponding normalized eigenvectors
$\bhe_s(\bfk)$ are the phonon polarization vectors.
In the long wavelength limit,
phonons have a linear dispersion relation 
\begin{align}
\omega_s(\bfk) = c_s(\hbk) k + {\cal O}(k^2),
\end{align}
where $c_s(\hbk)$ is the sound speed of the phonon mode in propagation direction
$\hbk=\bfk/k$. Generally speaking, polarization vectors $\bhe_s(\hbk)$ are
neither parallel nor perpendicular to $\hbk$. However, for
$k a \lesssim 1$, two of the phonon modes in a cubic lattice
are approximately transverse, and the third mode is approximately 
longitudinal. 

In Fig.~\ref{fig:dispersion} we show the phonon dispersion relations
and polarization of a body centered cubic
(BCC) lattice which are calculated from the dynamic matrix.
The lower frequency modes in the left-panel of
Fig.~\ref{fig:dispersion} correspond to the modes that are mostly
transverse with  $\bhe_s\cdot \hbk\approx 0$, and the higher frequency
modes are mostly longitudinal with $|\bhe_s\cdot \hbk|\approx 1$.  

Despite the relatively large spread of velocities associated with the
transverse modes it is often useful to represent them by an ``average
velocity'' denoted as $c_t$. This is typically defined by taking the
limit of $k \rightarrow 0$ in which case    
\begin{align}
c_t = \frac{\alpha\wpl}{\kD}
\approx 0.0031 \pfrac{\alpha}{0.4} \pfrac{Z}{30}
\pfrac{A}{80}^{-2/3}\rho_{10}^{1/6}, 
\label{eq:ct}
\end{align}
where  $\alpha\approx0.39$ according to \citet{Chabrier:1992}, and
\begin{align}
\kD = (6\pi^2 n_I)^{1/3}
\approx (0.41 a)^{-1}
\end{align}
is the Debye wave number. An approximate relation for the
longitudinal mode is given by 
\begin{align}
\omega_l^2(k) =\frac{\wpl^2}{\epsilon(k)}
\approx \frac{\wpl^2}{1+(\kTF/k)^2}.
\label{eq:omega-l}
\end{align}

\begin{figure*}[htb] 
  \begin{center}
    $\begin{array}{@{}ccc@{}}
      \includegraphics*[scale=0.55]{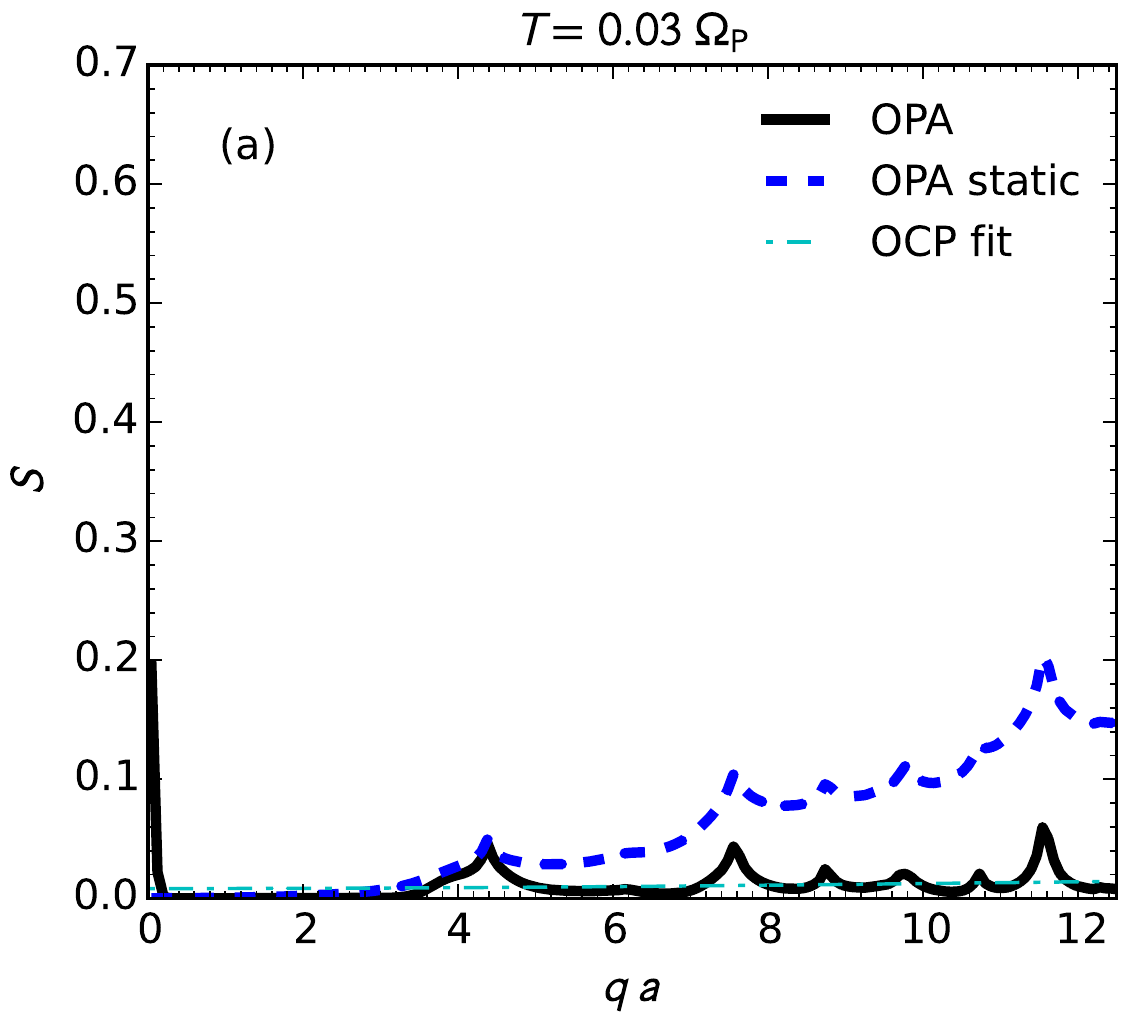} &
      \includegraphics*[scale=0.55]{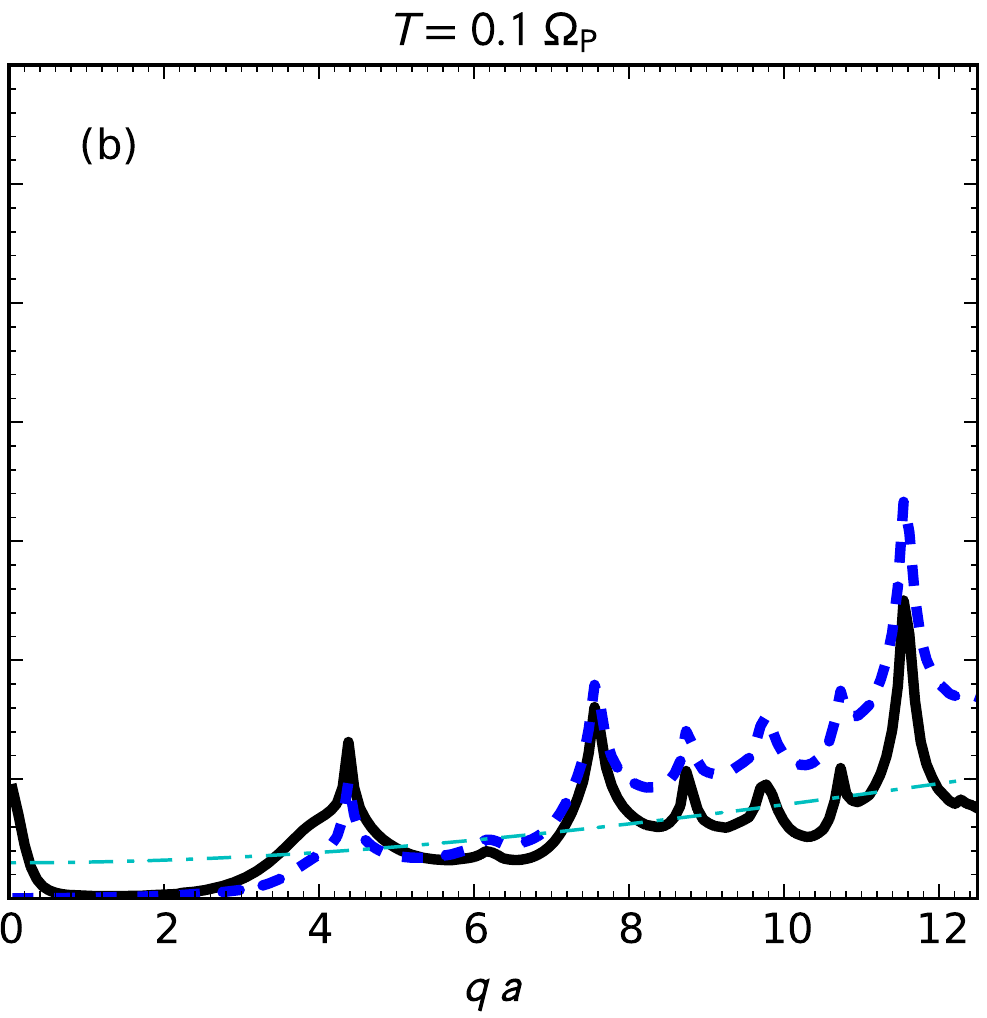} &
      \includegraphics*[scale=0.55]{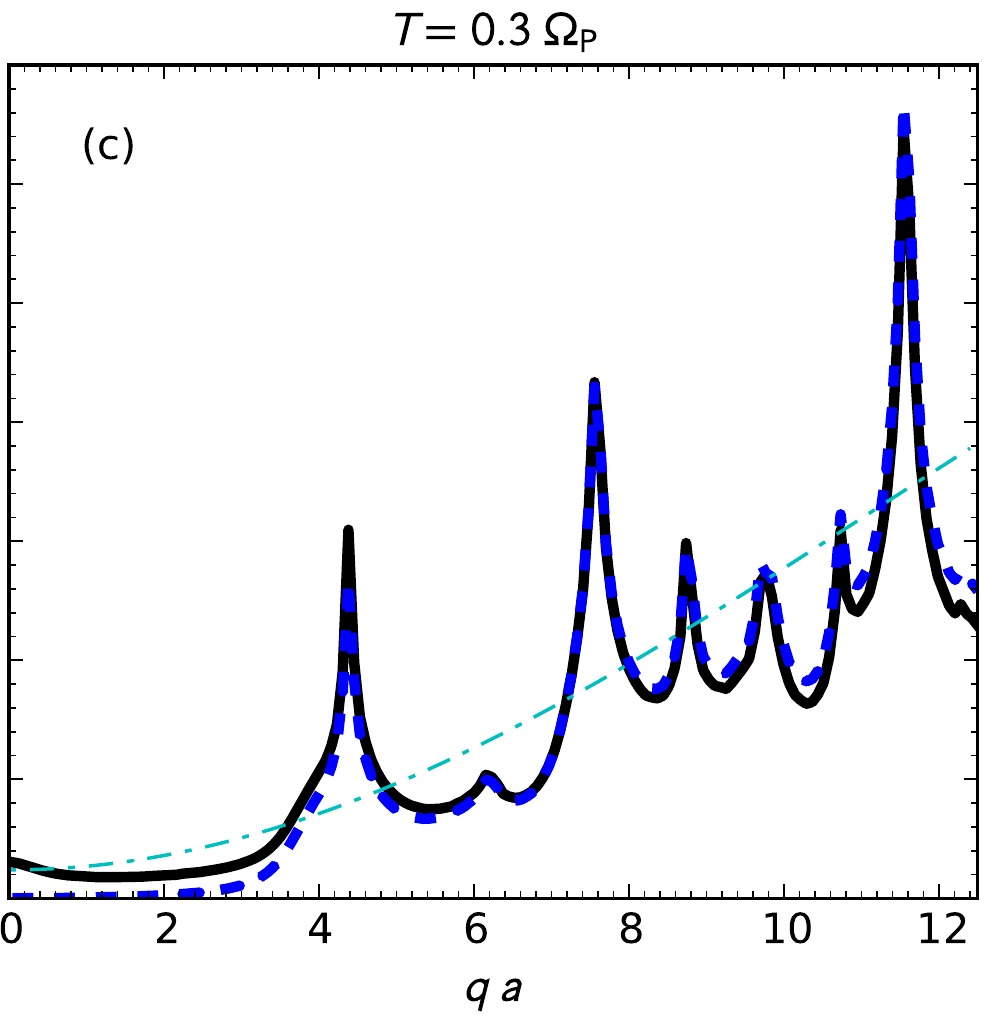}
      \end{array}$
\end{center}
  \caption{(Color online) The one-phonon approximation results of
    $S'_\kappa(q)$ (thick solid curves)    and
    $S'(q)$ (thick dashed curves)
    for the LD matter at the three temperatures as
    labelled. The results of $S'_\kappa(q)$ obtained using the fitting
    formula by \citet{Potekhin:1999}, which is based on the harmonic 
    approximation for the one-component Coulomb plasma and which includes
    multi-phonon contributions, are also
    shown (as dot-dashed curves) for comparison.}   
\label{fig:Skappa}
\end{figure*}

At low temperature it is useful to write the dynamic structure
function as a sum of the contributions from $n$-phonon processes
($n=0,1,\ldots$): 
\begin{align}
S(\omega,\bfq) = S^\p{0}(\omega,\bfq) + S^\p{1}(\omega,\bfq) + \cdots\,. 
\end{align}
The elastic or Bragg scattering is the 0-phonon contribution and is given by 
\begin{align}
S^\p{0}(\omega,\bfq) = e^{-2W(q)} \delta(\omega) N \sum_\bfK \delta_{\bfq,\bfK},
\end{align}
where $\bfK$ is a reciprocal lattice vector, and 
\begin{align}
e^{-2W(q)} = \exp\left(-\langle [\bfq\cdot\bfu(0)]^2 \rangle_T\right)
\end{align}
is the Debye-Waller factor which accounts for the suppression of
coherent scattering by thermal and  
quantum fluctuations of the ions. As mentioned earlier, the 0-phonon
contribution does not affect electron  
scattering.  At low temperature electron-ion scattering is dominated by the
1-phonon contribution
\begin{widetext}
\begin{align}
S^\p{1}(\omega,\bfq)
= \frac{e^{-2W}}{2M}\sum_{s,\bfK} 
\int\rmd^3k~
 \frac{[\bfq\cdot\bhe_s(\bfk)]^2}{\omega_s(\bfk)}
\delta^3(\bfK+\bfk-\bfq)
\left[\frac{\delta(\omega-\omega_s(\bfk))}{e^{\omega_s(\bfk)/T}-1}
+\frac{\delta(\omega+\omega_s(\bfk))}{1-e^{-\omega_s(\bfk)/T}}\right],
\label{eq:S1}
\end{align}
\end{widetext}
where the phonon momentum $\bfk$ is restricted to the first Brillouin
zone \cite{FlowersItoh:1976,Ashcroft}. In this case $S'_\kappa(q)$ in
Eq.~\eqref{eq:nu_kappa} can be replaced by
\begin{align}
  S^\OPA_\kappa(q)
  =\int_{-\infty}^\infty\rmd\omega\,  \avg{S^\p{1}(\omega,\bfq)}_\hbq
  w_\kappa(\omega/T, q).
\end{align}

Note that large-angle scattering
involves a finite  $|\bfK| \gg |\bfk|$ where the crystal absorbs a large
component of the momentum. This is well-known as the Umklapp process in
solid state physics \cite{Ashcroft}.  \citet{FlowersItoh:1976} realized
that 
these processes dominate over normal process (with $\bfK=0$)
in the neutron star context for typical temperatures of interest
because $\pF\gg\kD$. However, transitions with small $\bfk$ and finite
$\bfK$ are suppressed due to electronic band structure effects which
we shall now briefly discuss.

Although it is generally a good approximation to assume that electrons
are free, on patches of the electron Fermi surface which intersect
with the Brillouin zone boundaries, the effect of the periodic
background ion potential is large. It distorts the Fermi surface and
creates a band gap  in the electron
spectrum at the Fermi surface which is given by 
\begin{equation} 
\Delta\varepsilon(\pF) \simeq
\frac{4e^2}{3\pi}~\frac{\mathrm{e}^{-W(\pF)}~F(\pF)}
{\epsilon(\pF)}~\pF 
\end{equation} 
where $F(\pF) $ is the charge form factor of the nucleus \cite{Pethick:1996yj}. 
This gap can suppress the Umklapp processes when 
 \begin{equation} 
 T\lesssim T_\mathrm{U}\simeq c_t~\Delta\varepsilon(\pF) ,
\label{eq:T_um}
 \end{equation} 
\cite{Ziman:1960,RaikhYakovlev:1982,Chugunov:2012}.  
From Eq.~\eqref{eq:T_um} we can deduce that $T_\mathrm{U} < 10^{-2} \wpl$.
In what follows we restrict our analysis to the regime where $T$ is in the
range $10^{-2}-1~\wpl$ where the effects  
due to the band gap in the electron spectrum can be safely neglected
\cite{Chugunov:2012}. 


To determine the temperature regimes where the static approximation
$S'_\kappa(q)\approx S'(q)$ is valid, we compute the static structure
function $S'(q)$ in OPA:
\begin{align}
  S^\OPA(q) = \int_{-\infty}^\infty \rmd\omega\, 
  \avg{S^\p{1}(\omega, \bfq)}_\hbq.
  \label{eq:sofq-OPA}
\end{align}
In Fig.~\ref{fig:Skappa} we compare $S^\OPA_\kappa(q)$ and $S^\OPA(q)$
of the LD matter at three different temperatures, $T/\wpl=0.03$,
$0.1$ and $0.3$, respectively. 
The right panel of this figure shows
that, even at $T=0.3~\wpl$,  
$S'_\kappa(q)= S'(q)$ is already a good approximation.
The left panel of Fig.~\ref{fig:Skappa} shows that, at low temperature
$T= 0.03~\wpl$, the exponential factor
$1/(e^{\omega/T}-1)$ in Eq.~\eqref{eq:w} dominates, and
$S'_\kappa(q)\ll S'(q)$ in most of the range of $q$.
The middle panel of Fig.~\ref{fig:Skappa} with $T=0.1~\wpl$
illustrates the competition between two
factors in the expression of $w_\kappa$ at moderate temperatures,
which are discussed earlier following
Eq.~\eqref{eq:sofq} in Section~\ref{sec:kappa}.
For large-angle scattering with $q a \gtrsim 5$ the exponential factor
still dominates, and $S'_\kappa(q) < S'(q)$.
For small-angle
scattering with $q a \lesssim 5$, however, the factor $\pF^2/q^2$
dominates, and $S'_\kappa(q)>S'(q)$.  
We note that at low temperature $S'_\kappa(q)$ has a peak at
$q\rightarrow 0$ because 
\[w_\kappa(\omega/T\ll1,q a\ll1)\approx 1+\frac{3}{\pi^2}\pfrac{\pF}{T}^2
\pfrac{\omega}{q}^2,\]
and the second term in this expression always dominates under the typical
conditions in the neutron star crust.

For comparison we also show in Fig.~\ref{fig:Skappa} the results of
$S'_\kappa(q)$ 
obtained using the fitting formula by \citet{Potekhin:1999}
which is based on the harmonic 
approximation for the one-component Coulomb plasma and which includes 
multi-phonon contributions.

\section{Static structure function and Monte Carlo Simulations}
\label{sec:mc}

The neutron star crust spans the regimes where
purely classical simulations are sufficient and where quantum effects
start to play a significant role.  We can use classical and Quantum
Monte Carlo simulations (CMC and QMC)
to address these conditions. The QMC simulations have the classical
simulations as a specific limit.  Both the CMC and QMC calculations
can easily address the static structure function $S(q)$. They can also
be used to compute further information about the energy dependence  of
the response as well as other observables.

In CMC the kinetic and potential energies
are independent variables. Hence the positions of the particles can be
sampled independently of their momentum. We use a simple version of
the Metropolis Monte Carlo method to sample the positions of the
nuclei in periodic boundary conditions at fixed density and
temperature. 
The simulations use $N \gtrsim 1000$ particles, initially at
predetermined lattice sites in a periodic cubic box with length $L =
(N /n_I)^{1/3}$.
Proposed particle moves $\{\bfx_i\}\rightarrow\{\bfx'_i\}$ have
equal transition probabilities as their reverses:
\begin{align}
  \caT ( \{\bfx_i\} \rightarrow \{\bfx'_i\})
  &= \caT ( \{\bfx'_i\} \rightarrow  \{\bfx_i\}),
\end{align}
and they are accepted with probabilities
\begin{align}
  P ( \{\bfx_i\} \rightarrow \{\bfx'_i\})
  = \left\{\begin{array}{ll}
  e^{- \Delta E/T_\sigma} & \text{ if } \Delta E \geq 0, \\
  1 & \text{ otherwise.}
  \end{array}\right.
  \label{eq:P}
\end{align}
For CMC $T_\sigma=T$, and the energy change is the same as the change
in total potential energy:
\begin{align}
  \Delta E = E_\text{pot}(\{\bfx'_i\}) -
  E_\text{pot}(\{\bfx_i\}),
\end{align}
where
\begin{equation}
  E_\text{pot}(\{\bfx_i\})  =  \sum_{i<j} V (| \bfx_i - \bfx_j |).
  \label{eq:V}
\end{equation}
In Eq.~\eqref{eq:V}
the sums over $i$ and $j$ run over the particles
in the simulation volume plus their periodic images.  Typically plus or minus
one image in each direction (i.e.\ 27 periodic boxes in total)
is sufficient in these simulations because of the screening of the ion-ion
potential.
Standard Ewald summation is also possible but would be slower.
Detailed balance ensures that the Markov chain constructed
with the method described above
will converge eventually to sample particle positions
proportional to the partition function.

Quantum fluctuations become important when
$T/\wpl\lesssim 1$.
For such scenarios we used path integral 
QMC simulations (see, e.g., \cite{Ceperley:1995}).
The single position for
each particle in the classical simulation becomes a path in path
integral simulations with periodic
boundary conditions in imaginary time. Boson or fermion path integrals
would require us to include exchanges in the imaginary time boundary
conditions with the appropriate statistics, e.g.\ $-1$ for odd permutations
of fermions.
Because the characteristic distance $\lambda_I$ of ion motion in lattice
is much shorter than the inter-ion distance $a$,
quantum statistics (the boson or fermion nature of nuclei) is not important,
and we can consider the particles as distinguishable. 

In (path integral) QMC simulations the imaginary time or inverse
temperature $\beta=1/T$ 
is split into $N_\beta$ slices. 
Each slice is a
classical $N$-particle system described above but with effective
temperature 
\begin{align}
T_\sigma = (\Delta \tau)^{-1} = \left(\frac{\beta}{N_\beta}\right)^{-1}.
\end{align}
For QMC each imaginary-time slice involves a high-temperature expansion of the
propagator $\exp ( -H \Delta\tau)$. For a large enough number of
slices $N_\beta$ 
the results are independent of the number of slices.
Typically of order 10 slices are required in the present calculations.

As in CMC, the Markov chain is again constructed
by moving the particles according to the acceptance probability
defined in Eq.~\eqref{eq:P}. For QMC the energy change includes the
changes in both kinetic and potential energies:
\begin{align}
  \Delta E &= [E_\text{pot}(\{\bfx'_{i,\sigma}\}) +
    E_\text{kin}(\{\bfx'_{i,\sigma}\})]
  \nonumber\\
  &\quad -
      [E_\text{pot}(\{\bfx_{i,\sigma}\}) +
        E_\text{kin}(\{\bfx_{i,\sigma}\})],
\end{align}
where
\begin{align}
  E_\text{kin}(\{\bfx_{i,\sigma}\}) &= \sum_{i=1}^N\sum_{\sigma=1}^{N_\beta}
  \frac{(\bfx_{i,\sigma+1} - \bfx_{i,\sigma})^2}{2 M (\Delta\tau)^2},\\
  E_\text{pot}(\{\bfx_i,\sigma\})  &= \sum_{\sigma=1}^{N_\beta}
  \sum_{i<j} V (| \bfx_{i,\sigma} - \bfx_{j,\sigma} |)
\end{align}
with $\bfx_{i,N_\beta+1} = \bfx_{i,1}$.
Clearly, a CMC simulation can be considered as a special case of
QMC simulation with $N_\beta=1$.

\begin{figure*}[htb] 
\begin{center}
$\begin{array}{@{}cc@{}}
\includegraphics*[scale=0.55]{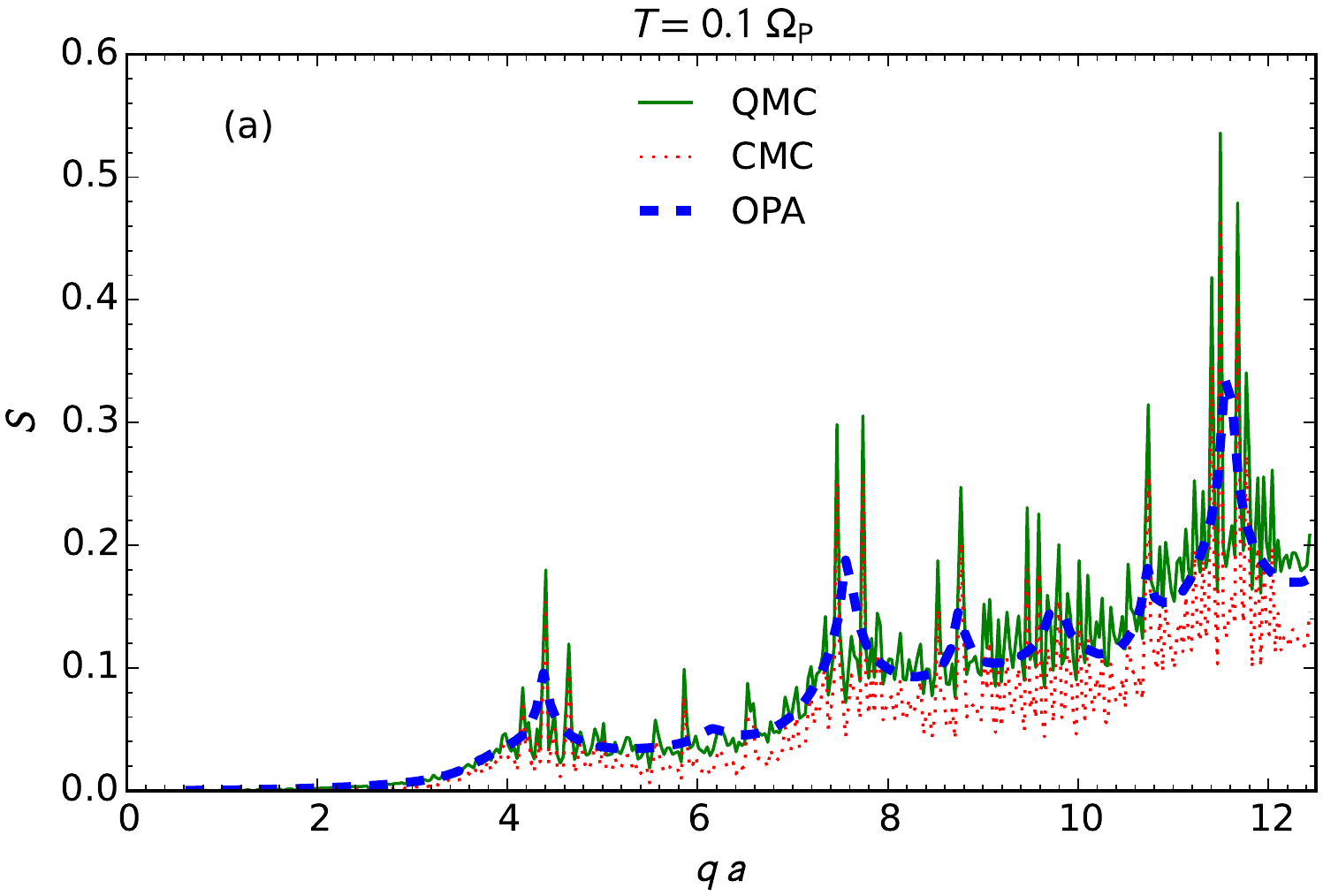} &
\includegraphics*[scale=0.55]{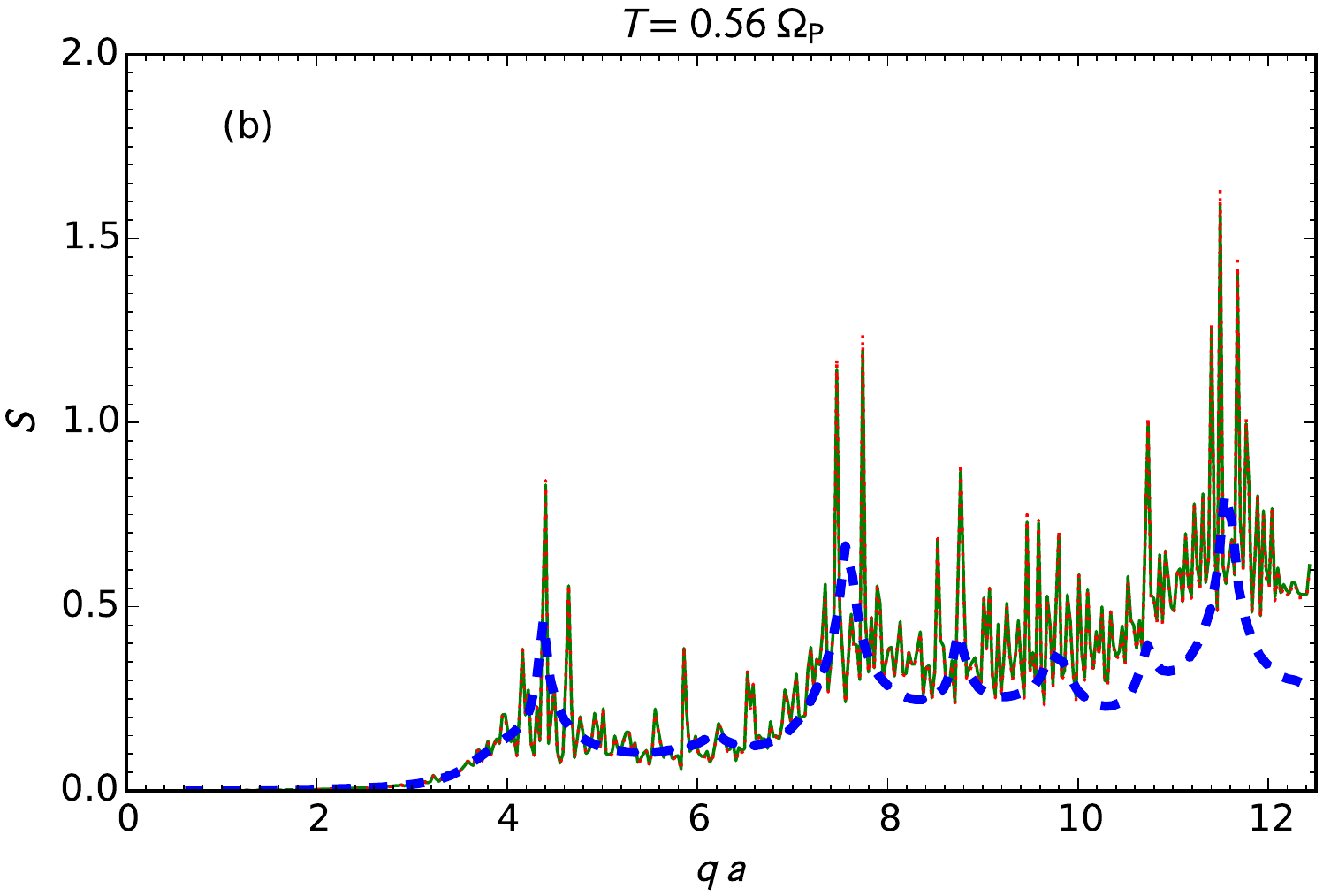}
\end{array}$
\end{center}
\caption{(Color online) The inelastic part of static structure function $S'(q)$
  of the LD matter at the two temperatures as labelled.
  The results are obtained using the one-phonon approximation (thick
  dashed curves), and quantum and classical Monte Carlo simulations
  (solid and dotted curves), respectively.} 
\label{fig:S}
\end{figure*}

The static structure function is then obtained from the points sampled
after convergence. In Monte Carlo simulations
\begin{align}
S(q) 
&=  \frac{1}{N N_\beta}\left\langle 
\sum_{\sigma=1}^{N_\beta} \sum_{i,j=1}^N 
e^{\rmi\bfq\cdot(\bfx_{i,\sigma} - \bfx_{j,\sigma})} \right\rangle_{\hat{\bfq},T},
\label{eq:S-QMC}
\end{align}
which includes both the one-phonon and multi-phonon contributions.
Because of the periodic condition,
\begin{align}
\bfq = \frac{2\pi}{L} (n_x\hbx + n_y\hby + n_z \hbz)
\end{align}
take discrete values,
where $n_{x (y,z)}$ are integers. 
To obtain the inelastic part
of the static structure function $S'(q)$ we simply remove
the points that correspond to the 
Bragg peaks in the BCC lattice. 
Other detailed structures predicted by QMC and CMC, i.e.\ the
smaller peaks and troughs away from the Bragg peaks (see
Fig.~\ref{fig:S}), are finite-size
artifacts whose amplitude decreases with increasing particle number in
the simulation.
However, the integrated strength over any reasonable interval in $q$ is
physically relevant and is insensitive to finite-size effects after the
numerical convergence has been achieved.

At high temperature $T\gtrsim\wpl$ all phonon modes are excited,
$S^\CMC(q)$ and $S^\QMC(q)$, which are
$S'(q)$ obtained using CMC and QMC simulations, respectively,
should agree. But at low
temperature $T\ll\wpl$ quantum fluctuations become prominent, and
$S^\QMC(q)>S^\CMC(q)$. 
In Fig.~\ref{fig:S} we compare $S^\CMC(q)$ and $S^\QMC(q)$ 
for the LD matter at two different temperatures,
$T/\wpl=0.1$ and $0.56$, respectively.
Indeed, $S^\QMC(q)$ is clearly larger than $S^\CMC(q)$ at
$T=0.1~\wpl$, but it is  
somewhat surprising that $S^\CMC(q)$
and $S^\QMC(q)$ agree very well even at temperature as low as
$T=0.56~\wpl$.

For comparison we also show $S^\OPA(q)$ in Fig.~\ref{fig:S}.
This figure shows that
$S^\QMC(q)$ and $S^\OPA(q)$ agree with each other at low
$T$ and/or small $q$, although there exist rapid oscillations
in Monte Carlo results because of the finite size of the system.
At high $T$ and/or large $q$ multi-phonon contributions are signfiant, and the
one-phonon approximation breaks down.

Note that in Eq.~\eqref{eq:S-QMC} only the equal-time correlator
has been evaluated.  By including an offset in the imaginary times
between the evaluations of positions of particles $i$ and $j$,
\begin{equation}
S(s\Delta\tau, q) =  \frac{1}{N N_\beta}\left\langle 
\sum_{\sigma=1}^{N_\beta} \sum_{i,j=1}^N 
e^{\rmi\bfq\cdot(\bfx_{i,\sigma} - \bfx_{j,\sigma+s})}
\right\rangle_{\hat{\bfq},T},
\label{eq:Stq}
\end{equation}
one can obtain information about the energy dependence of the response
\cite{Ceperley:1995}.
It is also possible to calculate the properties of MCP in both the
classical and quantum regimes.  This would require 
simulations significantly larger than the OCP
studied here, to ensure that the periodic boundary conditions do not
impact the results.  Simulations of this magnitude should be readily achievable
on modern parallel computers.

\section{Results and discussion}
\label{sec:results}

\begin{figure*}[ht]
\begin{center}
$\begin{array}{@{}cc@{}}
\includegraphics*[scale=0.55]{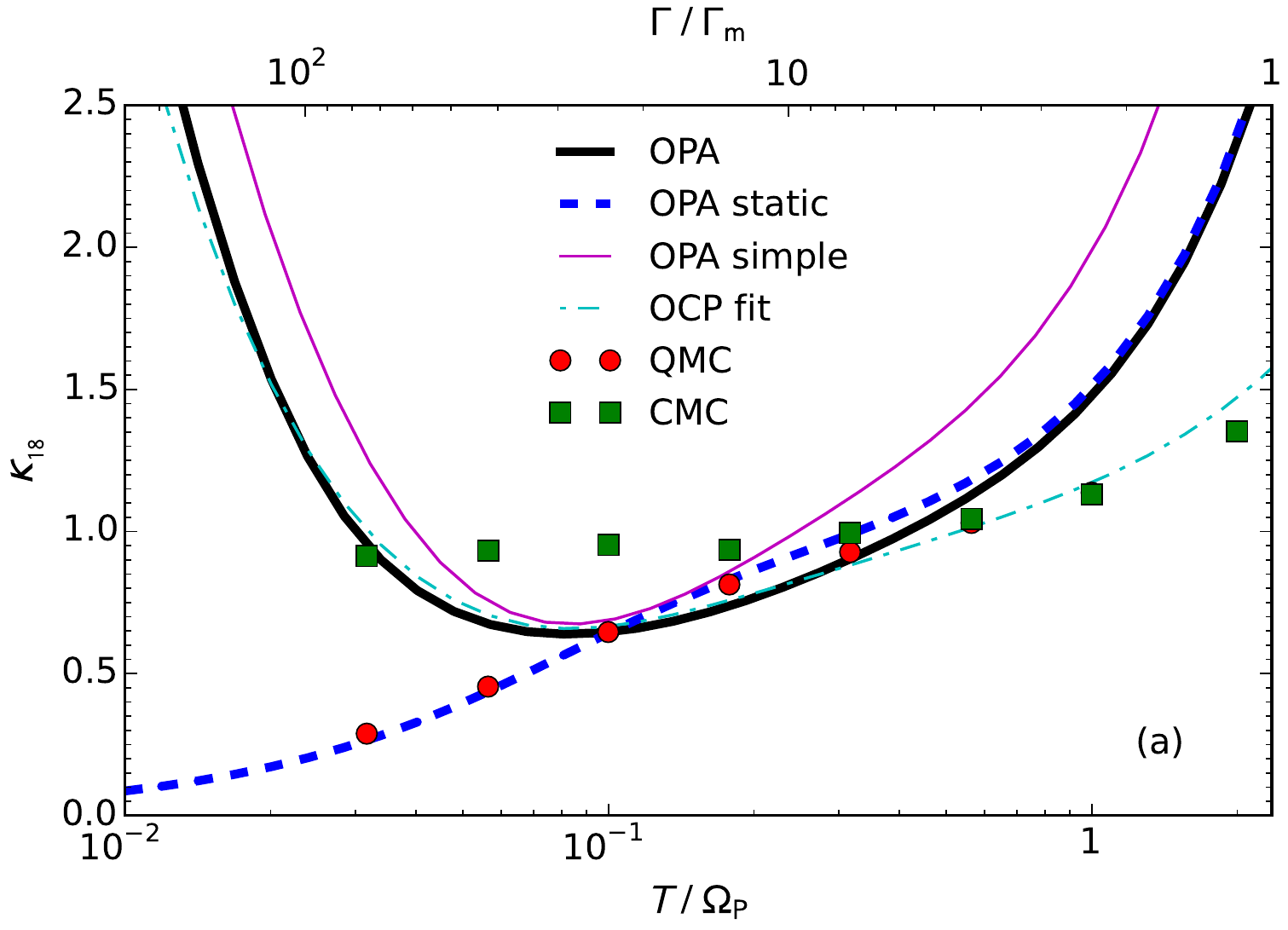}&
\includegraphics*[scale=0.55]{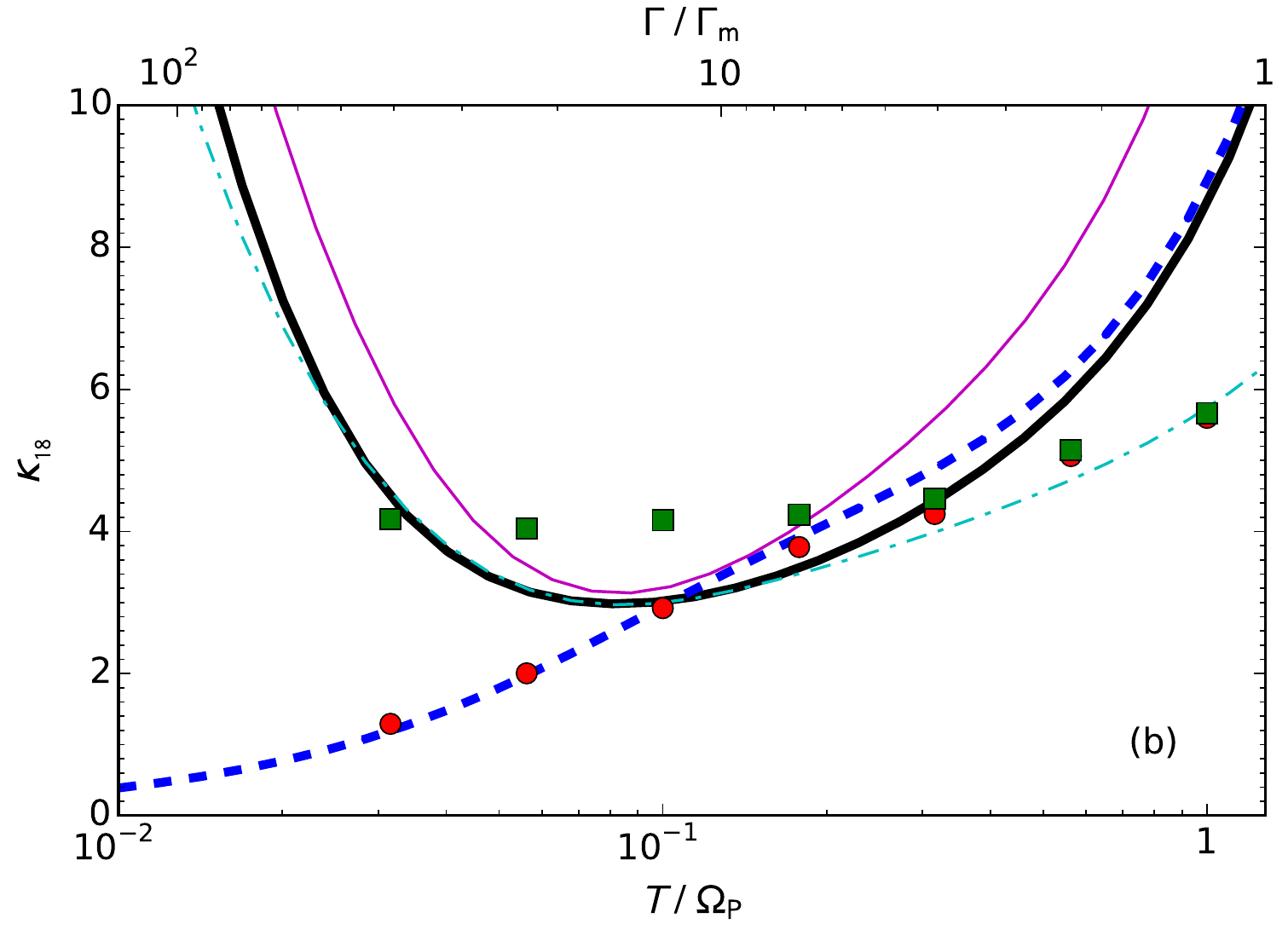}
\end{array}$
\end{center}
\caption{(Color online) Thermal conductivity of the LD (left panel) and HD matter
  (right panel) in units of
  $10^{18} \text{ erg cm}^{-1}\text{s}^{-1}\text{K}^{-1}$ and as a function
  of temperature.
  The results are obtained by replacing $S'_\kappa(q)$ in
  Eq.~\eqref{eq:nu_kappa} with $S^\OPA_\kappa(q)$ (thick
  solid curves), $S^\OPA(q)$ (thick dashed curves),
  $S^\OPA_\kappa(q)$ with approximate phonon dispersion relations [see
  Eqs.~\eqref{eq:ct} and \eqref{eq:omega-l}] (thin solid curves),
  fitting formula of $S'_\kappa(q)$ for one-component Coulomb plasma
  based on the harmonic approximation \cite{Potekhin:1999} (thin dot-dashed
  curves), $S^\QMC(q)$ (filled circles) and $S^\CMC(q)$ (filled
  squares), respectively.   
  \label{fig:kappa}}
\end{figure*}

We have calculated  the thermal conductivity of OCP for the LD and HD
ambient conditions outlined in Table~\ref{tab:cases} for the
temperatures of interest to neutron star phenomenology. Our primary
motivation to study the simple one component system was to obtain a
quantitative understanding of the effects of dynamical information and
quantum fluctuations at intermediate temperatures in the range
$0.1~\wpl \lesssim T \lesssim \wpl$. To this end we use the various
approximate methods outlined in previous sections to calculate
$S'_\kappa(q)$, which is the kernel function for computing effective
electron collision rate $\nu_\kappa$ in Eq.~\eqref{eq:nu_kappa}. 
We then calculated $\kappa$ for the
catalyzed neutron star matter with densities $10^{10}~\gcc$ (LD) and
$10^{11}~\gcc$ (HD), respectively.
The results are shown in Fig.~\ref{fig:kappa} where the thermal
conductivity is obtained by replacing $S'_\kappa(q)$ with
$S^\OPA_\kappa(q)$ (thick solid curves),
$S^\OPA(q)$ (thick dashed curves),
$S^\OPA_\kappa(q)$ with simple phonon dispersion relations [see
  Eqs.~\eqref{eq:ct} and \eqref{eq:omega-l}] (thin solid curves),
the fitting formula of $S'_\kappa(q)$ based on the harmonic
approximation \cite{Potekhin:1999} (thin dot-dashed
curves),
$S^\QMC(q)$ (filled circles)
and $S^\CMC(q)$ (filled squares), respectively.

A careful comparison of the results obtained using different
approximations for the function $S'_\kappa(q)$ provides the following
useful insights:  
\begin{enumerate}
\item  We find that it is adequate to set
  $S'_\kappa(q)= S'(q)$ in calculating $\kappa$ at 
  temperature as low as $T \approx  0.1~\wpl$.  A comparison between
  the thick dashed curves obtained using $S^\OPA(q)$ and the thick solid
  curves obtain using $S^\OPA_\kappa(q)$ supports this conclusion. The
  validity of this approximation is expected and well known for
  $T \gtrsim \wpl$.  Our results show that $S'_\kappa(q)\approx S'(q)$
  even at $T=0.3~\wpl$ (see Fig.~\ref{fig:Skappa}). Our results also show that,
  for $0.1  \lesssim T/\wpl \lesssim 0.3$, this
  approximate method can still be used to compute $\kappa$ even though
  $S'_\kappa(q)$ and $S'(q)$ differ. This is because there are
  two competing factors in
  $w_\kappa(\omega/T, q)$ which are discussed earlier following
  Eq.~\eqref{eq:sofq-OPA} in section \ref{sec:kappa}.  

\item Multi-phonon effects become relevant for
  $\Gamma/\Gamma_\melt \lesssim 4$ in our simulations. At higher
  $\Gamma$ or lower temperature
  the one-phonon approximation
  is adequate for OCP but is sensitive to the phonon dispersion relation. This
  is evident when we compare the thick solid curves, the thin
  solid curves and thin dot-dashed curves, which are obtained using the
  exact phonon dispersion relations from the dynamical matrix, the
  approximate phonon dispersion relations, and the harmonic
  approximation method with multi-phonon contributions
  \cite{Potekhin:1999},  respectively.  

\item The comparison between the results obtained using CMC and QMC simulations,
  shown by the filled squares and circles, respectively, indicates
  that quantum effects in thermal conductivity are significant when
  $T\lesssim  0.3~\wpl$ where classical calculations systematically
  underestimate $S'(q)$.  At $T \approx  0.1~\wpl$ CMC results overestimate
  $\kappa$ by $40-50\%$.  

\item The fitting formula for $S'_\kappa(q)$ based on harmonic
  approximation  \cite{Potekhin:1999} (dot-dashed curves) works quite
  well for the one-component Coulomb lattice 
  in the whole temperature range which we have studied. This can be seen
  when we compare them with the thick solid curves at
  $T\lesssim 0.3~\wpl$ and filled circles/squares at $T\gtrsim 0.3~\wpl$
  which are obtained using the one-phonon approximation and Monte
  Carlo simulations, respectively. At the highest temperature where 
  $ \Gamma \approx \Gamma_m$ the results obtained using the
  harmonic approximation include multi-phonon  
  excitations and agree well with the QMC results. This indicates that
  anharmonic effects are small even in this regime.  
  At the lowest temperatures, although the $S'_\kappa(q)$ obtained
  from the fitting formula based on the harmonic approximation differs
  from that obtained in the OPA 
  (see Fig.~\ref{fig:Skappa}),   
  the predictions for the thermal conductivity agree well as already discussed 
  in \cite{Potekhin:1999}.  
\end{enumerate}

Some of the trends emerging from these comparisons could have been be expected qualitatively. As previously alluded to,  
this systematic quantitative comparisons between QMC results and those obtained using the standard electron-phonon treatment 
provide a basis to asses the viability of using QMC calculations of $S(q)$ at low temperature for complex multi-component systems.  
In the standard treatment, multi-component systems are modeled as a regular lattice plus uncorrelated impurities, and electron 
scattering is assumed to arise due to incoherent contributions from electron-phonon and electron-impurity scattering. 
This treatment fails when the spatial distribution of the minority species is correlated, and QMC is a viable technique to calculate the role of these correlations in strongly coupled plasmas with $\Gamma\gg1$ in the regime when $T<\wpl$.

Our results show that $S(q)$ obtained from QMC is adequate to calculate $\kappa$ for $T\gtrsim 0.1~\wpl$, and  CMC may be adequate to compute thermal conductivity of OCP at $T\gtrsim 0.3~\wpl$. For lower temperatures, more detailed information about the energy dependence of the response is needed and we have briefly commented on how we can accesses this in the discussion following Eq.~\eqref{eq:Stq}. With a modest increase in computing resources, we have found that QMC can be used  to study large multi-component plasmas, and we are currently pursuing
calculations of $\kappa$  for ambient conditions encountered in accreting neutron stars. These
results will be reported elsewhere.

\begin{acknowledgments}
This collaborative work is supported in part by the DOE Topical Collaboration on
Neutrinos and Nucleosynthesis in Hot and Dense Matter. H.D.~thanks
LANL and NMC for providing the start-up funding and the computing
resources. S.R.\ was  
supported by the U.S.\ Department of Energy under Contract
No.\ DE-FG02-00ER41132.  
\end{acknowledgments}

\appendix

\bibliography{ref_noarxiv.bib}

\end{document}